\documentclass[doublecol]{epl2}

\usepackage{graphicx,color,amsmath,amssymb}

\newcommand{\bc}{\begin{center}}
\newcommand{\ec}{\end{center}}

\title{Unification   of    Bell,   Leggett-Garg   and   Kochen-Specker
  inequalities: Hybrid spatio-temporal inequalities}
\shorttitle{Unification of correlation inequalities} 
  
\author{Siddhartha Das\inst{1}\thanks{\email{sdas@students.iiserpune.ac.in}} \and
S. Aravinda\inst{2} \thanks{\email{aru@poornaprajna.org}}
\and R. Srikanth\inst{2,3}\thanks{\email{srik@poornaprajna.org}}
\and Dipankar Home\inst{4}\thanks{\email{dhome@jcbose.ac.in}}}
\shortauthor{S. Das \etal}
\institute{
	\inst{1} Indian Institute of Science Education and Research, Pune-
	411008, India \\
	\inst{2} Poornaprajna Institute of Scientific Research,
	Sadashivnagar, Bengaluru- 560080, India \\
	\inst{3} Raman Research Institute, Sadashivnagar,
	Bengaluru- 560060, India \\
	\inst{4} CAPSS, Bose Institute, Kolkata - 700 091, India
}


\abstract{The  Bell-type (spatial), Kochen-Specker  (contextuality) or
  Leggett-Garg  (temporal)  inequalities   are  based  on  classically
  plausible  but otherwise  quite  distinct assumptions.   For any  of
  these   inequalities,  satisfaction   is  equivalent   to   a  joint
  probability  distribution  for all  observables  in the  experiment.
  This implies a joint distribution  for all pairs of observables, and
  is indifferent to  whether or not they commute  in the theory.  This
  indifference underpins a unification  of the above inequalities into
  a general framework of  correlation inequalities.  When the physical
  scenario is such  that the correlated pairs are  all compatible, the
  resulting  correlation  is  non-signaling,  which may  be  local  or
  multi-particle,   corresponding   to   contextuality  or   Bell-type
  inequalities.   If   the  pairs  are   incompatible,  the  resulting
  correlation  corresponds to  Leggett-Garg  (LG) inequalities.   That
  quantum mechanics  (QM) violates  all these inequalities  suggests a
  close connection between the  local, spatial and temporal properties
  of the theory.   As a concrete manifestation of  the unification, we
  extend  the method  due to  Roy and  Singh (J.   Phys.  A,  11, L167
  (1978)) to  derive and study  a new class of  hybrid spatio-temporal
  inequalities, where the correlated  pairs in the experiment are both
  compatible or incompatible. The implications for cryptography
  and  monogamy inequalities  of the  unification are  briefly touched
  upon.}

\begin{document}
\maketitle

\section{Introduction}

Three  types of statistical  \textit{correlation inequalities}  may be
discerned  that probe  the nonclassical  aspects of  quantum mechanics
(QM), namely:  Bell-type or spatial  inequalities \cite{chsh69,bell+},
which   test    the   assumption   of    local-realism   \cite{epr35},
Kochen-Specker     \cite{ks67}    or     contextuality    inequalities
\cite{kcbs08,cab96,yuoh12}  that   test  noncontextual  realism,  and
finally    Leggett-Garg    or    temporal    inequality    \cite{lg85,
  lap06,kofbruk12,   ushalg,  lgrev},   that  test   macro-realism  or
noninvasive-realism.   Classical   theory  presumes  \textit{realism},
i.e., observations  merely reveal a  pre-existing value of  a measured
property.   The  quantum  mechanical  violation of  these  correlation
inequalities implies  that realism must sometimes be  given up, unless
classically  counter-intuitive  modes of  influence  are posited  (For
example, an explanation of the violation of Bell-type inequality would
require signaling over space-like separated events).

In  the  Bell-type and  contextuality  inequalities, the  experimental
\textit{correlata} (the experimentally correlated pair of observables)
are  mutually compatible  (or commuting,  in  QM), whereas  in the  LG
inequality, sometimes considered as the temporal analog of a Bell-type
inequality,   the  correlata   are  incompatible,   necessitating  the
measurements  to be  sequential.   Further, we  may  regard Bell  type
inequalities as a special  case of contextuality inequalities, wherein
context  is  provided  through  spatial  separation.   All  the  above
inequalities  have been  tested, modulo  certain arcane  loopholes, in
laboratories worldwide: for  Bell-type inequalites \cite{asp81, asp82,
  wei98,   sch10},  for  contextuality   inequalities  \cite{contexp1,
  contexp2}   and  LG   inequalities  in   both   macroscopic  systems
\cite{lgfried,lgwal}  and   microscopic  systems  \cite{gog10,  fed11,
  lgath11, maheshlg}.

The problem of unifying these inequalities into a single framework was
first considered  by Markiewicz et  al.  \cite{3uni}, who  showed that
the  existence of  a  joint probability  distribution  is the  general
assumption  underlying  all  three  types  of  inequalities.   As  one
consequence,  they demonstrated how  one type  of inequality  could be
transformed  into   another.   In   this  work,  we   examine  another
consequence of  the unification,  namely that of  hybrid inequalities,
which mix  correlations of  different types. Here  we make use  of the
interesting method due to Roy and Singh \cite{rs78,rs79}, who were the
first to derive from the local-realist condition testable inequalities
\textit{different} from the  Bell-CHSH-type inequalities. We adapt the
Roy-Singh  (RS)  method of  deriving  spatial  inequalities to  derive
contextual,    temporal   and    finally    hybrid   (spatio-temporal)
inequalities.  The  new  inequalities  we  obtain  are  no  harder  to
implement experimentally than the other inequalities.

The  remaining part  of this  article is  divided as  follows.  In the
following Section, we introduce the RS method for deriving
Bell-type inequalities.  Later,  we point out that the  existence of a
joint   distribution   underlies   these   inequalities,   essentially
revisiting a theorem  due to Fine \cite{fine82} in  the context of the
RS inequalities. That this observation also extends to the case
of contextuality and LG-type  inequalities is shown subsequently. As a
demonstration  of  the  generality  of  this result,  we  next  derive
contextuality and LG-type inequalities  via the RS method, followed by
a  novel  class  of   hybrid  spatio-temporal  inequalities.  After  a
discussion on monogamy inequalities, we present conclusions in the last section
\ref{sec:conclu}, touching on  some cryptographic ramifications of our
work.

\section{RS inequalities\label{sec:rs}}

The Roy-Singh  method provides an  elegant method to  derive local-realist
inequalities \cite{rs78}.   Suppose two  qubits in an  entangled state
are intercepted by two measuring devices geographically separated from
each other.  The first  device randomly measures property either $X_1,
X_2, \cdots $ on particle 1, and the other device either property $Y_1
,  Y_2  ,  \cdots  $  on particle  2.   Experimentally,  one  measures
bi-partite correlations  of the type $P(x_j,y_k)$, where  $x_j = \pm1$
and $y_k=\pm1$  are measurement outcomes of measuring  $X_j$ and $Y_k$
respectively.  The  assumption of local-realism (LR)  entails that for
each such  pair of  these variables, there  is a  deterministic hidden
variable (DHV) theory whereby
\begin{equation}
\langle X_jY_k \rangle = \int_\lambda d\lambda \rho(\lambda) X_j(\lambda)
Y_k(\lambda),
\label{eq:RS-}
\end{equation}
where $\lambda$ is a  ``complete'' or ``dispersion-free'' specification of
the   state   described   by   underlying   probability   distribution
$\rho(\lambda)$.  Roy and Singh \cite{rs78} consider quantities of the
form
\begin{widetext}
\begin{eqnarray}
&& \left(X_1^{(1)} \pm X_2 ^{(1)}\pm \cdots X_{m_1}^{(1)}+
Y_1^{(1)} \pm Y_2^{(1)} \pm \cdots \pm Y_{n_1}^{(1)}\right)^2 +
\left(X_1^{(2)} \pm X_2^{(2)} \pm \cdots X_{m_2}^{(2)} + \right. \nonumber \\
&&\left. 
Y_1^{(2)} \pm Y_2^{(2)} \pm \cdots \pm Y_{n_2}^{(2)}\right)^2 +
\cdots 
+
\left(X_1^{(q)} \pm X_2^{(q)} \pm \cdots X_{m_q}^{(q)} +
Y_1^{(q)} \pm Y_2^{(q)} \pm \cdots \pm Y_{n_q}^{(q)}\right)^2  
\ge q,
\label{eq:rsstart}
\end{eqnarray}
\end{widetext}
\begin{center}
\mbox{\textit{see eq.~\eqref{eq:rsstart}}}
\end{center}
where  $m_j+n_j  =  $  odd,  and  the  self-correlation  terms,  i.e.,
correlations between the  same particle, are so arranged  as to cancel
out.  Here  $X^{(j)}_k \in \{X_1,X_2,\cdots,X_n\}$  and $Y^{(j)}_k \in
\{Y_1,Y_2,\cdots,Y_m\}$, where $m, n$  are positive integers and $X_j,
Y_k \in \{\pm1\}$ . As a particular example, we consider
\begin{equation}
(X_1 - Y_1 - Y_2)^2 + (X_2 - Y_1 + Y_2)^2 \ge 2.
\label{eq:rs1}
\end{equation}
Expanding the  l.h.s of (\ref{eq:rs1}),  and using the  DHV assumption
Eq. (\ref{eq:RS-}), we obtain:
\begin{equation}
\langle \mathcal  {B} \rangle \equiv \langle X_1Y_1  \rangle + \langle
X_1Y_2 \rangle +  \langle X_2Y_1 \rangle - \langle  X_2Y_2 \rangle \le
2,
\label{eq:rsbell}
\end{equation}
which  just is  the  CHSH inequality  \cite{chsh69}.   By varying  the
odd-termed expressions in Eq.  (\ref{eq:rs1}),  in such a way that the
same-particle  correlations (self-correlations)  cancel  out, Roy  and
Singh obtain other Bell-type inequalities \cite{rs78}.

\section{Joint realism\label{sec:jr}}

In the above derivation of a Bell-type inequality, the following point
is   worth   noting:   that   in   the   step   represented   by   Eq.
(\ref{eq:rsstart}),  one presumes \textit{joint  realism} (JR)  of all
variables  in  the  experiment,   i.e.,  that  there  exists  a  joint
probability                      distribution                     (JD)
$P(x_1,x_2,\cdots,x_n,y_1,y_2,\cdots,y_m)$  of  all  variables in  the
experiment.   This  is  equivalent   to  the  DHV  considered  in  Eq.
(\ref{eq:RS-}) \cite{fine82}.  To see this, we note that the existence
of a DHV defines a JD
\begin{widetext}
\begin{equation}
P(x_1,\cdots,x_n,y_1,\cdots, y_m) = \int_\lambda \rho(\lambda)d\lambda
\tilde{X}_1(x_1,\lambda)\tilde{X}_2(x_2,\lambda)\cdots
\tilde{X}_n(x_n,\lambda)
\tilde{Y}_1(y_1,\lambda)\tilde{Y}_2(y_2,\lambda)\cdots
\tilde{Y}_m(y_m,\lambda),
\label{eq:RS}
\end{equation}
\end{widetext}
\begin{center}
\mbox{\textit{see eq.~\eqref{eq:RS}}}
\end{center}
where $\tilde{X}_j(x_j,\lambda)$  is an indicator  function that takes
the value  1 if  $X_j =  x_j$ given $\lambda$,  else 0,  similarly for
$\tilde{Y}_k(y_k,\lambda)$.  This  JD corresponds to  the two-particle
correlations of Eq. (\ref{eq:RS-}) since
\begin{eqnarray}
\langle   X_jY_k  \rangle   &\equiv&  \sum_{x_1\cdots,y_1,\cdots=\pm1}
P(x_1,x_2,\cdots,x_n,y_1,y_2,\cdots,y_m)x_j   y_k  \nonumber   \\  &=&
\sum_{x_j,y_k}  P(x_j,y_k)   x_jy_k  \nonumber  \\   &=&  \int_\lambda
\rho(\lambda)          \sum_{x_j,y_j}         \tilde{X}_j(x_j,\lambda)
\tilde{Y}_k(y_k,\lambda) x_jy_k
\label{eq:corr}
\end{eqnarray}
from which the rhs of  Eq.  (\ref{eq:RS-}) follows since by definition
$\sum_{x_j} \hat{X}_j(x_j,\lambda)x_j  = X_j(\lambda)$ and $\sum_{y_k}
\hat{Y}_k(y_k,\lambda)y_k  =  Y_k(\lambda)$.   That  JR  implies  that
Bell-type  inequalities hold  follows from  the previous  Section. The
converse of these  two arguments can be given  along the lines adopted
in  Ref   \cite{fine82},  thereby  establishing   the  equivalence  of
existence  of   JD,  of  DHV  and  the   satisfaction  of  correlation
inequalities.

\section{Unification\label{sec:uni}}

A  key  observation  is   that  the  satisfication  of  a  correlation
inequality  implies  that  consistent  JDs  exist  for  all  commuting
\textit{and  non-commuting} pairs of  observables.  The  proof follows
from the fact that the required JDs, say $P(x_1,y_2)$, can be obtained
by     tracing     out     all     the    other     variables     from
$P(x_1,\cdots,x_n,y_1,\cdots,y_m)$.  Suppose $S =  A \cup B$, where $A
\equiv \{X_j\}$ and  $B \equiv \{Y_k\}$ are two  groups of properties,
and  the experimental  outcome is  a correlation  of the  form $P(x_j,
y_k)$. The \textit{instantiation} of  the grouping does not matter: in
other words,  $A$ and  $B$ may refer  to two  geographically separated
particles,  or to two  different times  on the  same particle,  and so
forth.

From the perspective of an experiment, there are two cases to consider:

\textit{Compatible correlata.} That all elements within $A$ are mutually
  incompatible, as are all  elements within $B$, whereas every element
  $X_j$ is  compatible with  every element $Y_k$.   This compatibility
  means that the correlation $P(x_j,y_k)$ will be non-signaling, since
  measurements in  $A$ reveal no  information about operations  in $B$
  and vice versa.  A correlation inequality obtained here  may thus be
  referred to as  \textit{non-signaling}.  
  measured cross-correlations are  between compatible observables, and
  the cross-correlated observables have  pairwise JD, the violation of
  a   correlation  inequality   in   this  case   arises  from   local
  incompatibility  within $A$  and/or within  $B$.  This  explains the
  close   connection    between   non-commutativity   \cite{mj07}   or
  complementarity \cite{ghocomp} or Heisenberg uncertainty \cite{js10}
  on the one hand and nonlocality on the other.

\textit{Incompatible correlata.}  That elements  between $A$ and $B$, in
  addition   to   having   intra-group  incompatibility,   also   have
  \textit{inter-group}  incompatibility.    When  $A$  is   a  set  of
  measurements  on a  particle at  time  $t_A$, and  $B$ is  a set  of
  measurements on  the same particle at  time $t_B>t_A$.  Correlations
  are obtained  by \textit{sequential} measurements on  account of the
  incompatability.   The correlation  $P(x_j,y_k)$ will  be signaling,
  since the  outcome of $Y_k$ will  depend on which  $X_j$ preceded it
  \cite{arvsri}.   Here we are  led to  \textit{signaling} correlation
  inequalities. The term  signal is used in the  formal sense that the
  no-signaling  condition is violated,  and does  not refer  to signal
  between  spatially  seperated parties.   By  virtue of  relativistic
  causality,     signaling    correlations    cannot     pertain    to
  spacelike-separated   measurements   on   geographically   separated
  particles but they can  describe sequential measurements on the same
  particle.

Clearly,   Bell-type  and  contextuality   inequalities  are   of  the
non-signaling kind, while the temporal inequalities are necessarily of
the signaling kind.  Our observation  at the beginning of this Section
implies that a correlation inequality is indifferent as to whether the
correlation is signaling or otherwise:  in both cases, violation of an
inequality  means that  a  HDV  or equivalent  JD  description is  not
possible.   This indifference means  that we  can regard  the spatial,
temporal  and contextual  inequalities  as instantiations  of the  same
unified mathematical object, which is a HDV or a JD.

The  physical significance  of  this object  depends  on the  physical
scenario  at  hand.  Let  $A$  and  $B$  correspond to  geographically
separate  observers.   Assuming JR,  a  violation  of the  inequality,
implies a microscopic  spacelike signal in the sense  of $B$'s outcome
depending on  $A$'s setting or  vice versa.  Thus satisfaction  of the
inequality  is tantamount  to  the assumption  of local-realism  (LR).
Analogously, if $A$ and $B$ correspond to the same particle or system,
then  JD is  equivalent to  the assumption  of  non-contextual realism
(NCR).   On the other  hand, in  the case  of a  signaling correlation
inequality, $A$ and $B$ necessarily  pertain to the same particle.  In
this case JD requires  that subsequent measurements were not disturbed
by  the  earlier  ones,  and  thus tantamount  to  the  assumption  of
non-invasive realism (NIR) or macro-realism.

There are profound ramifications of this unification.  One is that the
method  to derive  any one  type of  inequality can  be  translated in
straightforward   fashion  to   yield  another   type   of  inequality
\cite{3uni}.   Accordingly,   we  adapt   the  RS  method   to  derive
contextuality and LG-type inequalities.   Second is that, in all three
types of inequalties , the  \textit{classical bound}, which is the rhs
(having value 2 in Eq.   (\ref{eq:rsbell}) ) is indifferent to whether
a  spatial,  temporal or  contextual  scenario  is instantiated.   The
Tsirelson bound  is also identical for the  considered cases involving
qubits,  though in general  this bound  can be  different \cite{3uni}.
Another  ramification is that  QM violates  these inequalities  in all
three instantiations,  meaning that its  nonclassical structure exists
under spatial, temporal and  contextual instantiations, showing a deep
connection between  these aspects.  One  can in principle  imagine toy
theories where this three-fold nonclassicality need not hold true, for
example,  classical  theory,  genaralized  local  theory  \cite{bar07}
equipped with a temporal PR box for sequential measurements.

As a final ramification of the unification, we demonstrate below using
the  RS  method \textit{hybrid  inequalities}  in  which  some of  the
correlata are compatible and some  are not. Such an inequality will be
a    simultaneous   test    of   say    NIR   and    LR,    which   is
noninvasive-local-realism (NILR).  One  can similarly instatiate JR to
obtain an  inequality that tests  NCR and LR  together, or NIR  and LR
together or all three together. A spatio-temporal hybrid inequality is
considered in detail later below.

\section{Contextuality and temporal inequalities from the RS method
\label{sec:gen}}

The generality of the observation that JR underlies all the three
types of inequalities is seen by adapting the RS method to
derive spatial and contextuality inequalities.

As a prelude  to deriving the hybrid inequality via  the RS method, we
discuss how one can obtain contextuality and LG-type inequalities.  It
is  clear  that  the  RS  procedure  already  yields  state  dependent
contextuality  inequalities  when  the  labels  $X$  and  $Y$  of  the
Bell-type inequalities are interpreted as defining pairwise compatible
properties of the same particle.

The contextuality inequality  studied by Ref. \cite{kcbs08} can be
obtained starting from:
\begin{equation}
(X_{1}+X_{2}+X_{3})^{2}+(X_{3}+X_{4}+X_{5})^{2}+(X_{1}-X_{3}+X_{5})^{2}\geq 3
\end{equation}
which yields:
\begin{equation}
X_{1}X_{2}+X_{2}X_{3}+X_{3}X_{4}+X_{4}X_{5}+X_{5}X_{1}\geq -3,
\end{equation}
where the consecutive subscript indices refer to commuting observables
of the same  particle.  This can be generalized  in similar fashion to
give
\begin{widetext}
\begin{equation}
X_{1}X_{2}+X_{2}X_{3}+X_{3}X_{4}+X_{4}X_{5}+X_{5}X_{6}+X_{6}X_{7}+X_{7}X_{1}+5\geq0
\label{eq:this1}
\end{equation}
\end{widetext}
\begin{center}
\mbox{\textit{see eq.~\eqref{eq:this1}}}
\end{center}
starting from
\begin{widetext}
\begin{equation}
(X_{1}+X_{2}+X_{3})^{2}+(X_{3}+X_{4}+X_{5})^{2}+(X_{5}+X_{6}+X_{7})^{2}+(X_{1}-X_{3}+X_{5})^{2}+(X_{1}-X_{5}+X_{7})^{2}+5\geq0
\label{eq:this2}
\end{equation}
\end{widetext}
\begin{center}
\mbox{\textit{see eq.~\eqref{eq:this2}}}
\end{center}
and applying the RS method. Similarly for an odd $n$-polygon,
we obtain chained inequalities like
\begin{widetext}
\begin{equation}
X_{1}X_{2}+X_{2}X_{3}+X_{3}X_{4}+X_{4}X_{5}+\cdots+ X_{n-1}X_{n}+X_{n}X_{1}+n-2\geq0.
\label{eq:this3}
\end{equation}
\end{widetext}
\begin{center}
\mbox{\textit{see eq.~\eqref{eq:this3}}}
\end{center}
For example, starting from
\begin{equation}
(X_{1}-X_{2}+X_{3})^{2}+(X_{3}-X_{4}+X_{5})^{2}+(X_{1}-X_{3}+X_{5})^{2}\geq 3
\end{equation}
one obtains    via    the    RS   method,    the    inequality    $
X_{1}X_{2}+X_{2}X_{3}+X_{3}X_{4}+X_{4}X_{5}-X_{5}X_{1}\geq  3.  $ This
kind of chained inequality  \cite{BC90} can be readily generalized for an
odd   $n$-polygon,   to   obtain   inequalities   of   the   type:   $
X_{1}X_{2}+X_{2}X_{3}+X_{3}X_{4}+X_{4}X_{5}+                     \cdots
+X_{n-1}X_{n}-X_{n}X_{1}\geq  n-2.  $  To obtain  LG-type inequalities
via  the RS  procedure,  we  drop all  particle  labels. For  example,
writing
\begin{equation}
(L-K-M)^2 + (J -K + M)^2 \ge 2,
\label{eq:rs2}
\end{equation}
we  obtain via  the RS  method the  LG-type inequality:  $  \langle JK
\rangle + \langle KL \rangle + \langle LM \rangle - \langle JM \rangle
\le 2,  $ which  can be experimentally  tested on qubits  by measuring the
properties $J, K$ or $L$ at time $t_1$, and $K, L$ or $M$ at time $t_2 >
t_1$.  However, because $\langle KL\rangle = \langle LK\rangle$ by JR,
the above inequality is equivalent to the usual LG inequality.

\section{Hybrid inequalities\label{sec:rshybrid}}

If we retain the particle labels  of the original RS procedure, but do
not require the \textit{self-correlation} terms to be elimintated, the
result  is \textit{hybrid}  inequalities  that encompass  correlations
paired  without  restriction.   Self-correlation  terms  $\langle  X_j
X_k\rangle$  or $\langle Y_j  Y_k\rangle$ correspond  to incompatible,
sequential    measurements,   assumed    to    satisfy   NIR,    while
cross-correlation  terms   $\langle  X_j  Y_k\rangle$   correspond  to
compatible measurements  assumed to satisfy LR.   The resulting hybrid
spatio-temporal inequality  is thus a  joint test of \mbox{LR  + NIR},
i.e., local-noninvasive-realism (LNIR).

As an example, starting from
\begin{equation}
(X_2 - X_1 + Y_1)^2 + (X_1 - Y_2 + Y_1)^2 \ge 2
\label{eq:ekh}
\end{equation}
we obtain via the RS method, 
\begin{equation}
F \equiv  \langle X_1X_2 \rangle +  \langle X_1 Y_2  \rangle - \langle
X_2Y_1\rangle + \langle Y_1Y_2\rangle \le 2.
\label{eq:eqh}
\end{equation}
The experimental protocol
 can be as follows: at $t_1$,  Alice measures $X_1$
or not, and Bob measures $Y_1$  or not; at $t_2$, Alice measures $X_2$
or not, and  Bob measures $Y_2$ or not. This gives rise to the following
16 possibilities
$$
\{\emptyset_X, X_1, X_2, X_1X_2\} \times
\{\emptyset_Y, Y_1, Y_2, Y_1Y_2\}
$$  where $\emptyset_X$  $(\emptyset_Y)$ denotes  a nonaction  by $A$
$(B)$ while $X_1X_2$ or $Y_1Y_2$  is a pair of sequential measurements
by a party  on the respective particle.  Of these 16,  only 9 turn out
to be experimentally  admissible, while 3 yield two  data points.  The
observational  set corresponding to  the pair  $(\emptyset_X, Y_1Y_2)$
yields   the   data  $\langle   Y_1Y_2\rangle$,   as  does   $(X_1X_2,
\emptyset_Y)$ yield $\langle X_1  X_2\rangle$.  The pairs $(X_1, Y_2)$
and $(X_2,Y_1)$ yield, respectively, the data $\langle X_1 Y_2\rangle$
and $\langle X_2Y_1\rangle$.  The pair $(X_1, Y_1Y_2)$ yields the data
$\langle    X_1Y_1\rangle$   (which   does    not   appear    in   the
Ineq. (\ref{eq:eqh}))  and $\langle Y_1Y_2 \rangle $  but not $\langle
X_1Y_2\rangle$  since  measurement of  $Y_1$  is  invasive for  $Y_2$.
Similarly,  the  pair  $(X_1X_2,Y_1)$  contributes the  data  $\langle
X_1X_2\rangle$. On  the other hand, $(X_2,Y_1Y_2)$  yields, apart from
$\langle X_2Y_1\rangle$ also  $\langle Y_1 Y_2\rangle$. Similarly, the
pair  $(X_1X_2, Y_2)$ contributes  data $\langle  X_1X_2\rangle$ and
$\langle  X_1Y_2\rangle$,  while  $(X_1X_2, Y_1Y_2)$  yields  $\langle
X_1X_2 \rangle$ and $\langle Y_1 Y_2\rangle$.

For the  Bell state $|\Psi^-\rangle  = \frac{1}{\sqrt{2}}(|01\rangle -
|10\rangle)$,   we   find   that   $F  =   \hat{x}_1\cdot\hat{x_2}   +
\hat{x}_1\cdot\hat{y_2}        -       \hat{x}_2\cdot\hat{y_1}       +
\hat{y}_1\cdot\hat{y_2}$ which attains a  value of $2\sqrt{2}$ for the
settings  where $\hat{x}_2,  \hat{x}_1,\hat{y}_2,  \hat{y}_1 $  differ
sequentially  by  $\pi/4$.   This  is  in  fact  the  Tsirelson  bound
\cite{cir80}   for  this  inequality.    By  the   unification,  Ineq.
(\ref{eq:eqh}) is  indifferent to whether the correlata  belong to the
same particle  or two different particles.  Thus,  we can re-interpret
$X_1, Y_1$  as belonging to one  particle (say $A$) and  $X_2, Y_2$ as
belonging  to   another  particle  (say   $B$).   As  a   result,  the
self-correlation    terms     in    Ineq.     (\ref{eq:eqh})    become
cross-correlations,  while the cross-correlation  terms remain  so, in
effect  converting  the   spatio-temporal  inequality  into  the  CHSH
inequality \cite{chsh69}, where  all correlata are cross-correlations.
Now we  make use of the  fact that these  cross-correlation terms when
evaluated   for  a   singlet,  are   numerically  the   same   as  the
(state-independent)        temporal        self-correlation       term
\cite{fritz10}. Thus, the Tsirelson bound for Ineq.  (\ref{eq:eqh}) is
the  same as  that  for  the CHSH  inequality,  which is  $2\sqrt{2}$,
attained for  singlets. In general, for arbitrary  spin, the Tsirelson
bound may be different under the conversion \cite{3uni}.

An interesting  fact here  is that product  states can  violate hybrid
inequality (\ref{eq:eqh}).   This is because the hybrid inequality
can be  violated from  contributions both from  the local  sector (via
quantum  invasiveness)  and nonlocal  sector,  and  the former,  being
state-independent,  can be  made sufficiently  high for  the violation
even  with separable states.   Let  the two  particles be  in the
product      state     $|\psi_A\rangle|\psi_B\rangle      $,     where
$|\psi_A\rangle\langle\psi_A|         =         \frac{1}{2}(1        +
\hat{n}_A\cdot\vec{\sigma})$   and   $|\psi_B\rangle\langle\psi_B|   =
\frac{1}{2}(1  + \hat{n}_B\cdot\vec{\sigma})$. Then  we find  for this
state that
\begin{equation}
F =  \hat{x}_1\cdot\hat{x}_2 + \left(\hat{x}_1\cdot\hat{n}_A\right)
\left(\hat{y}_2\cdot\hat{n}_B\right)                               -
\left(\hat{x}_2\cdot\hat{n}_A\right)\left(\hat{y}_1\cdot\hat{n}_B\right)
+ \hat{y}_1\cdot\hat{y}_2.
\end{equation}
For  the settings where  $\hat{x}_2, \hat{x}_1,\hat{y}_2,  \hat{y}_1 $
differ  sequentially   by  $\pi/4$,  and  $\hat{n}_A   =  \hat{n}_B  =
\hat{y}_2$,   we   find   that   Ineq.    (\ref{eq:eqh})   is   yields
$\frac{3}{\sqrt{2}} \approx 2.12$.

The  lhs in  Ineq.  (\ref{eq:eqh})  is  the expectation  value of  the
correlation  operator  $  \hat{F} =  [\textbf{X}_1\cdot\textbf{X}_2  +
  \textbf{Y}_1\cdot\textbf{Y}_2]   +  [(\textbf{X}_1\cdot\vec{\sigma})
  \otimes               (\textbf{Y}_2\cdot\vec{\sigma})              -
  (\textbf{X}_2\cdot\vec{\sigma})                               \otimes
  (\textbf{Y}_1\cdot\vec{\sigma})]  $  where  the  first term  in  the
square  brackets (denoted $S_1$)  is a  state-independent one  and the
second   term   (denoted    $S_2$)   is   the   state-dependent   one.
Interestingly,  we  find after  some  straightforward manipulation,  $
(S_2)^2     =    2[\mathbb{I}     -    (\textbf{X}_1\cdot\textbf{X}_2)
  (\textbf{Y}_1\cdot\textbf{Y}_2)                                     -
  ((\textbf{X}_1\times\textbf{X}_2)\cdot\vec{\sigma})           \otimes
  ((\textbf{Y}_1\times \textbf{Y}_2)\cdot\vec{\sigma})],
\label{eq:S2}
$
which is also  state-independent.  Putting together contributions from
$S_1$  and  $S_2$,  we  find  for this  state-independent  quantity  a
Tsirelson-like behavior in that
\begin{widetext}
\begin{eqnarray}
\left|\left|\sqrt{\hat{F}^2}\right|\right|                        &\le&
\left|\left|(\hat{x}_1\cdot\hat{x}_2)  +  (\hat{y}_1\cdot\hat{y}_2)  +
\sqrt{2}\sqrt{           \left\langle           (\mathbb{I}          -
  (\hat{x}_1\cdot\hat{x}_2)(\hat{y}_1\cdot\hat{y}_2)                  -
  ((\hat{x}_1\times\hat{x}_2)\cdot\vec{\sigma})                 \otimes
  ((\hat{y}_1\times\hat{y}_2)\cdot\vec{\sigma}
  )\right\rangle}\right|\right| \nonumber  \\ &=& \left|\cos\theta_1 +
\cos\theta_2   +\sqrt{2}\sqrt{1   -  (\cos(\theta_1)\cos(\theta_2)   +
  \sin(\theta_1)\sin(\theta_2))}\right|         \nonumber\\        &=&
\left|\cos\theta_1      +     \cos\theta_2      +\sqrt{2}\sqrt{1     -
  \cos(\theta_1-\theta_2)}\right| \le 2\sqrt{2},
\label{eq:tsirelson}
\end{eqnarray}
\end{widetext}

\begin{center}
\mbox{\textit{see eq.~\eqref{eq:tsirelson}}}
\end{center}

which  is the  same as  the  Tsirelson bound  for the  spatio-temporal
Ineq. (\ref{eq:eqh}).  This is saturated for example  with $\theta_1 =
-\theta_2 = \pi/4$.

\section{Monogamy inequalities}
Here we point out that the  RS formalism can be used to derive an
interesting monogamous  behaviour between  the violations of  the KCBS
and CHSH inequalities, which was  recently obtained by Kurzynski et al
\cite{KCK0}.    Such   a  behavior   generalizes   the  monogamy   for
non-signaling  spatial correlations  and  non-disturbing contextuality
correlations,  and   is  suggested  by  the  above   idea  of  hybrid
correlations.

Consider the RS requirement
$\{(X_{3} + Y_{1} + Y_{2})^2 + 
(X_{1} + Y_{1} - Y_{2})^2\} + 
\{(X_{1}+X_{2}+X_{3})^{2} + 
(X_{3}+X_{4}+X_{5})^{2} + 
(X_{1}-X_{3}+X_{5})^{2}\} \geq 5$, which
yields the monogamy relation as follows
\begin{eqnarray}
&& \langle X_{3}Y_{1}+X_{3}Y_{2}+X_{1}Y_{1}-X_{1}Y_{2}\rangle +
\langle X_{1}X_{2} + \nonumber \\
&& X_{3}X_{2}+ X_{3}X_{4}+X_{4}X_{5}+ X_{5}X_{1}\rangle \geq -5 \nonumber \\ 
&& \langle CHSH \rangle_{XY} + \langle KCBS \rangle_{X} \geq -5.
\label{eq:monors0}
\end{eqnarray}
Suppose  that  the  particle   $X$  is  at  least  3-dimensional,  and
$[X_j,X_k]=0$   iff  $|j-k|\le   1   \mod  5$.    To   see  that   Eq.
(\ref{eq:monors0}) is  a monogamy relation, i.e.,  one whose violation
leads to  signaling and  would thus be  disallowed, we  re-arrange the
terms   in  the   lhs  to   obtain  $\langle   X_3Y_2-X_1Y_2+X_1X_2  +
X_3X_2\rangle  +  \langle  X_3Y_1  +   Y_1X_1  +  X_5X_1  +  X_4X_5  +
X_3Y_4\rangle$, the former which has  a CHSH-like form, and the latter
a KCBS-like form.  Assuming  that no-signaling holds between particles
$X$ and $Y$, we  can write down a JD $p_C$ for  both the CHSH-like and
KCBS-like  terms,  such  that experimentally  accessible  correlations
$p_E$  are recovered.   For example,  for the  CHSH-like term,  in the
method  of Ref.  \cite{KCK0},  we can  guarantee  satisfaction of  the
inequality   by   constructing   a   JD   $p_C(x_1,x_3;   y_1,x_2)   =
(p_E(x_1,x_2,y_2)p_E(x_2,x_3,y_2))/p_E(x_2,y_2)$,              provided
$\sum_{x_1}p_E(x_1,x_2,y_2)     =     \sum_{x_3}p_E(x_2,x_3,y_2)     =
p_E(x_2,y_2)$,  which is  the  no-signaling,  or generally,  the
no-disturbance condition.

\section{Conclusion and discussions\label{sec:conclu}}

We  have  shown  that  the  experimental  violation  of  the  spatial,
contextual  and   temporal  correlation  inequalities   are  different
instantiations of the same underlying property of joint realism of all
observables  in   the  experiment.   As  one   manifestation  of  this
unification,  the classical  bound  is indifferent  to the  particular
instantiation.   As another  manifestation,  a new class
of hybrid  spatio-temporal
inequalities has been proposed, which  is shown to be maximally violated
even by product states.

The unification of the correlations provides a framework to understand
the  relation between  the  security of  the  three broadly  different
quantum   cryptography  protocols   \cite{chitra}:   those  based   on
nonlocality, such  as the Ekert protocol \cite{ekert},  those based on
conjugate coding  and invasive measurement, such  as BB84 \cite{bb84},
and   those   based   on   a   single  orthogonal   basis   like   the
Goldenberg-Vaidman  (GV)  protocol  \cite{gv,preeti} or
the counterfactual quantum protocols \cite{N09,XSS2}.  The  connection
between  JD and  security  is that  (non-signaling) correlations  that
violate  a Bell-type  inequality  possess cryptographically  desirable
properties like monogamy, no-cloning, uncertainty, incompatibility and
secrecy \cite{mag06}.  The  connection to JD is manifest  in the Ekert
protocol, where security requires violation of a Bell-type inequality.
The correlations that arise in the  BB84 does not lead to violation of
LG  inequality,  even  though   it  is  based  on  incompatibility  or
invasiveness of local measurements. It  is this lack of violation that
makes  BB84 insecure in  the device-independent  cryptography scenario
\cite{di}.  Likewise in the GV protocol, though the observed data does
not lead to data for which a correlation inequality can be tested, the
encoding  basis  is  incompatible  with  any basis  accessible  to  an
eavesdropper.

DH  and SA  acknowledge support  from DST  for Project  No.  SR/S2/PU-
16/2007 and the INSPIRE  fellowship IF120025, respectively.  DH and SA
acknowledge  the  Centre  for   Science,  Kolkata  and  Manipal  University,
respectively,  for  research  support.   The  authors  thank  the
Referee for valuable suggestions.

\bibliographystyle{eplbib}
\bibliography{quantuni}

\end{document}